\documentclass[fleqn,twoside]{article}
\usepackage{espcrc2}
\usepackage{graphicx,epsfig,rotating}

\newcommand{\lsim}   {\mathrel{\mathop{\kern 0pt \rlap
  {\raise.2ex\hbox{$<$}}}
  \lower.9ex\hbox{\kern-.190em $\sim$}}}
\newcommand{\gsim}   {\mathrel{\mathop{\kern 0pt \rlap
  {\raise.2ex\hbox{$>$}}}
\lower.9ex\hbox{\kern-.190em $\sim$}}}

\def\be{\begin{equation}}
\def\ee{\end{equation}}
\def\ba{\begin{eqnarray}}
\def\ea{\end{eqnarray}}
\def\eps{{\varepsilon}}

\title{Ultra High Energy Cosmic Rays Spectra in Top-Down models}

\author{R. Aloisio\address[LNGS]{INFN - Laboratori Nazionali del Gran Sasso,
        I--67010 Assergi (AQ), Italy} 
        \thanks{Talk presented by R. Aloisio},
        V. Berezinsky\addressmark[LNGS]
        and M. Kachelrie{\ss}\address{Max-Planck-Institut f\"ur Physik 
                  (Werner-Heisenberg-Institut), D--80805 M\"unchen, Germany}}
       
\begin{document}

\begin{abstract}

In this work we present a detailed computation of the  spectra of UHECR 
in the top-down scenario. We compare the spectra of hadrons obtained by 
two different methods in QCD and supersymmetric (SUSY) QCD with large primary 
energies $\sqrt{s}$ up to $10^{16}$ GeV. The two methods discussed are a 
Monte Carlo (MC) simulation and the evolution of the hadron fragmentation 
functions as described by the Dokshitzer-Gribov-Lipatov-Altarelli-Parisi 
(DGLAP) equations. The hadron spectra obtained by the two methods agree
fairly well in the interesting energy range $10^{-5}M_{X}<E<0.3M_{X}$ ($M_X$
is the energy scale of the process $M_{X}\ge 10^{12}$ GeV). We have also 
computed the spectra of photons, neutrinos and nucleons obtaining a good 
agreement with other published results. The consistency of the spectra 
computed by different methods allows us to consider the spectral shape as a 
signature of the production model for UHECR, such as the decay of super heavy
relic particles or topological defects.

\vspace{1pc}
\end{abstract}

\maketitle

\section{Introduction}
Ultra High Energy Cosmic Rays (UHECRs) are still an open problem in 
astro-particle physics. The 11 Akeno Grand Air Shower Array (AGASA) events 
with energy larger than $10^{20}$ eV \cite{expCR} contradict the expected 
suppression of the UHECR spectrum due to the interaction with the Cosmic 
Microwave Background (CMB) radiation, the Graisen-Zatsepin-Kuzmin (GZK) 
cut-off \cite{GZK}. On the other hand the HiRes data seems to be 
consistent with the GZK cut-off picture \cite{expCR}. If the UHECR primaries
are protons and if they propagate rectilinearly, as the claimed correlation 
with BL-Lacs at energy $4-8\times 10^{19}$ eV implies, than their 
sources must be seen in the direction of the highest energies events with 
energies up to $2-3 \times 10^{20}$ eV detected by HiRes, 
Fly's Eye and AGASA \cite{expCR}. At these energies the proton 
attenuation length is only about $20-30$ Mpc and no counterparts 
in any frequency band was observed in the direction of these UHECR events. 
This is a strong indication that CR particles with energies larger than 
$10^{20}$ eV may have a different origin from those with lower energies. 

Another important point of this discussion is related to the low energy
part of the CR spectrum, there are, infact, strong evidences that these CR 
with energies $1\times 10^{18} {\rm eV}\le E\ge 7-8 \times 10^{19} {\rm eV}$
are extragalactic protons, most probably from Active Galactic Nuclei (AGNs). 
This statement is based on some robust experimental and theoretical evidences:
(i) extensive air shower (EAS) data confirm protons as primaries, 
(ii) the dip seen with great accuracy in the 
data of AGASA, HiRes, Fly's Eye and Yakutsk, is a strong signature
of the propagation of UHE protons in the extragalactic space, (iii) the
beginning of the GZK cut-off seen in the spectra of AGASA and HiRes.
Excluding the correlation of UHECR with BL Lacs from the analysis becomes 
also possible the propagation of protons in very strong magnetic fields. 
Nevertheless, also in this case, the lack of a nearby source in the
direction of the highest energy events (e.g. at $E\sim 3\times
10^{20}$~eV) remains a problem. In fact, for very strong field strengths 
$B\sim 1$~nG: the deflection angle, $\theta \sim l_{\rm att}/r_H=
3.7^{\circ} B_{\rm nG}$ given by the attenuation length $l_{\rm att}$
and the Larmor radius $r_H$, is small and sources should be seen.

Many ideas have been put forward aiming to explain the observed
superGZK ($E \gsim (6-8)\times 10^{19}$~eV) events: strongly interacting 
neutrinos and new light hadrons as unabsorbed
signal carriers, $Z$-bursts, 
Lorentz-invariance violation, Topological Defects (TD)
and Superheavy Dark Matter (SHDM) (see \cite{Xpart} for reviews).
The two last models listed above, that represent the most promising top-down 
models, share a common feature: UHE particles are produced in the decay of 
superheavy (SH) particles or in their annihilation. In the case of TD they 
are unstable and in the case of SHDM  long-lived particles. 
We shall call them collectively $X$ particles. Annihilation 
takes place in the case of monopolonia, necklaces \cite{neckl} 
and SHDM particles within some special models. From the point of view of 
elementary particle physics all these processes proceed in a way similar to
$e^+e^-$~annihilation into hadrons: two or more off-mass-shell
quarks and gluons are produced and they initiate QCD cascades. 
Finally the partons are hadronized at the confinement radius. 
Most of the hadrons in the final state are pions and thus the typical
prediction of all these models is the dominance of photons at the
highest energies $E \gsim (6-8)\times 10^{19}$~eV. Let us now concentrate 
our attention to the computation of the UHECR spectrum produced in the 
decay of $X$ particles. 

The spectrum of hadrons produced in the decay/annihilation of $X$ particles 
is another signature of models with superheavy $X$ particles. 
The mass of the decaying particle, $M_X$, or the energy of annihilation
$\sqrt{s}$, is in the range $10^{13}$ -- $10^{16}$~GeV. 
The existing QCD MC codes become numerically unstable at much smaller
energies, e.g., at $M_X \sim 10^7$~GeV. Moreover, the computing time 
increases rapidly going to larger energies. In this work we will review our 
results obtained, in the computation of the top-down spectrum of UHECR, 
using two different computational techniques: one based on a MC computation 
scheme \cite{ABK} and the other based on the DGLAP evolution equations 
\cite{ABK}. In both cases SUSY is included in the computation.
Monte Carlo simulations are the most physical approach for high
energy calculations which allow to incorporate many important physical
features as the presence of SUSY partons in the cascade and
coherent branching. The perturbative part of our MC
simulation scheme is similar to other existing MC codes it also includes 
in a standard way SUSY and hence is reliable. 
For the non-perturbative hadronization part an original phenomenological 
approach is used in Ref.~\cite{ABK}. The fragmentation of a parton $i$ 
into an hadron $h$ is expressed through perturbative fragmentation function 
of partons $D_i^j(x,M_X)$, that represents the probability of fragmentation 
of a parton $i$ into a parton $j$ with momentum fraction $x=2p/M_X$, convoluted
with the hadronization functions $f_j^h(x,Q_0)$ at scale $Q_0$,
that is understood as the fragmentation function of the parton $i$ into the 
hadron $h$ at the hadronization scale $Q_0\simeq 1.4$ GeV \cite{ABK}. 
To obtain the fragmentation functions of hadrons one has:
\be
D_i^h(x,M_X)=
\sum_{j=q,g}\int_x^1\frac{dz}{z}D_i^j(\frac{x}{z},M_X)f_j^h(z,Q_0)
\label{hfunc}
\ee
where the hadronization functions do not depend on the scale
$M_X$. This important property of hadronization functions allows us to 
calculate $f_i^h(x,Q_0)$ from available LEP data, $D_i^h(x,M_X)$ at the 
scale $M_X=M_Z$, and then to use it for the calculation of fragmentation 
functions $D_i^h(x,M_X)$ at any arbitrary scale $M_X$. Our approach reduces 
the computing time compared to usual MC simulations and allows a fast 
calculation of hadron spectra for large $M_X$ up to $M_{\rm GUT}$. 

The perturbative part of the MC simulation in Ref.~\cite{ABK} includes
standard features such as  angular ordering, which provides the
coherent branching and the correct Sudakov form factors, as well as SUSY
partons. Taking into account SUSY partons results only in small
corrections to the production of hadrons, and therefore a simplified
spectrum of SUSY masses works with good accuracy.
The weak influence of supersymmetry is explained by the decay of SUSY 
partons, when the scale of the perturbative cascade reaches the SUSY scale
$Q_{\rm SUSY}^2 \sim 1~{\rm TeV}^2$. Most of the energy of SUSY partons
remains in the cascade in the form of energy of ordinary partons, left
after the decay of SUSY partons. The qualitatively new effect caused by 
supersymmetry is the effective  production of the Lightest Supersymmetric 
Particles (LSP), which could be neutralinos or gluinos.
The fragmentation functions $D_i^h(x,M_X)$ at a high scale $M_X$ can
be calculated also evolving them from a low scale, e.g. $M_X=M_Z$, where they 
are known experimentally or with great accuracy using the MC scheme. This 
evolution is described by the Dokshitzer-Gribov-Lipatov-Altarelli-Parisi 
(DGLAP) equation \cite{DGLAP} which can be written as
\be
\partial_t D_i^h=\sum_j\frac{\alpha_s(t)}{2\pi}P_{ij}(z)\otimes
D_j^h(x/z,t)\,,
\label{DGLAP-eq}
\ee
where $t=\ln(s/s_0)$ is the scale, $\otimes$ denotes the convolution
$f\otimes g=\int_z^1 dx/x f(x)g(x/z)$, and $P_{ij}$ is the splitting function 
which describes the emission of parton $j$ by parton $i$. 
Apart from the experimentally rather well determined quark fragmentation 
function $D_q^h(x,M_Z)$, also the gluon fragmentation function $D_g^h(x,M_Z)$ 
is needed for 
the evolution of Eq.~(\ref{DGLAP-eq}). The gluon FF can be taken either from MC
simulations or from fits to experimental data, in particular
to the longitudinal polarized $e^+e^-$ annihilation cross-section and
three-jet events.

The first application of the DGLAP method for the calculation of hadron spectra
from decaying superheavy particles has been made in Refs.~\cite{previous}.
The most detailed calculations have been performed by Barbot and Drees 
\cite{previous}, where more than 30 
different particles were allowed to be cascading and the mass spectrum of the
SUSY particles was taken into account. Although at $M_Z$, which is
normally the initial scale in the DGLAP method, the fragmentation
functions for supersymmetric partons are identically zero, they can be
calculated at larger scales $t$: 
SUSY partons are produced above their mass threshold, when their
splitting functions are included in Eq.~(\ref{DGLAP-eq}).
In \cite{ABK} we proved that this method is correct. Also, the LSP spectrum 
can be computed within the DGLAP approach \cite{previous}. 

In this paper we shall study the agreement of the two methods: MC
and DGLAP equations for the calculation of spectra produced in the decay
or annihilation of superheavy particles. We shall also
compare the results obtained by different groups comparing the
calculated spectra with recent ones measured by UHECR experiments.

The paper is organized as follows: in Section 2 we will compare the two 
computation schemes described, referring the reader to our paper \cite{ABK}
for more details. Photon, neutrino and proton spectra, needed for UHECR
calculations, are computed in Section 3 and compared with the spectra
obtained in \cite{previous}. In Section 4 we will 
discus the consequences of our results for models of SHDM and TD
in the explanation of the UHECR spectrum. We will conclude in section 5.

\section{MC and DGLAP comparison}

\begin{figure}
\epsfig{file=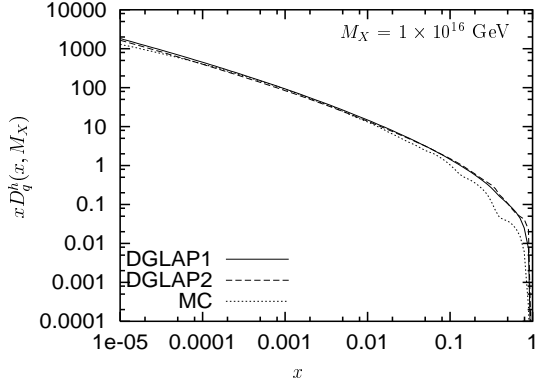,,height=5cm,width=7cm}
\caption{\label{fig1}
Comparison of SUSY DGLAP and SUSY MC fragmentation functions for
$M_X=1\times 10^{16}$ GeV with quark as a primary parton.
SUSY DGLAP FFs are calculated for 10 TeV as the starting scale
(solid line) and for $M_Z$ (broken line). 
SUSY MC FF is shown by dotted line.}
\end{figure}

In this Section we shall compare the hadron spectra computed by
the two methods discussed above, MC and DGLAP. 
In Figure \ref{fig1} we plot the FFs $D_i^h(x,M_X)$ calculated by SUSY MC and 
SUSY DGLAP methods for $M_X=1\times 10^{16}$ GeV and $i=q$. In the DGLAP
method the SUSY FFs have been evolved from the ones obtained with the
SUSY MC at the scale $\sqrt s= 10M_{\rm SUSY}\approx 10$~TeV. One can
see the good agreement between DGLAP (solid curve) and MC (dotted curve).
This good agreement holds also for other (lower) scales $M_X$ and for
other initial partons, e.g. gluon, squark or gluino. 
One can see that the MC and DGLAP spectra slightly differ at very low 
$x$ and have a more pronounced disagreement at large values of $x$.
The discrepancy at low $x$ is due to coherent branching, that is included 
in the MC scheme while it cannot be embedded in the DGLAP one.
At large $x$, the calculations by both methods suffer from uncertainties, 
particularly the MC simulation. In this region the results are sensitive to 
the details of the hadronization scheme (see, e.g., the problem of
HERWIG~\cite{previous} with the overproduction of protons at large $x$). 
\unitlength1.0cm
When one does not have the initial SUSY FFs from a MC simulation, the
question arises how to proceed. As was first suggested by Rubin
\cite{previous}, the initial FFs can be taken as the ones for ordinary
QCD at the low scale $\sqrt s=M_Z$, while the production of SUSY
partons is included in the splitting functions assuming
threshold behavior at $M_{\rm SUSY} \sim 1$~TeV. We can check this
assumption computing the SUSY FF in both ways. In Figure \ref{fig1}
we present the SUSY FFs $D_i^h(x,M_X)$ for $i=q$ and $M_X=1\times
10^{16}$ GeV, evolved from the initial scale $\sqrt s=M_Z$ (dashed
curve). The good agreement between the two DGLAP curves proves the
validity of the assumption made.

\section{Photon, neutrino and nucleon spectra}

The spectra of photons, neutrinos and nucleons produced by the decay of
superheavy particles are of practical interest in high energy
astrophysics. These spectra $D_i^a(x,M_X)$ with $a=\gamma,\nu,N$ can
be also considered as FFs. Because the dependence on the type $i$
of the primary parton is weak, we shall omit the index $i$ from now on,
keeping $a$ as subscript.

Till now we concentrated our discussion on the total number
of hadrons ($a=h$) described by the FF $D_h(x,M_X)$, 
but in fact we have performed similar calculations separately for
charged pions and protons+antiprotons. The procedure of the
calculations is identical to that already described for the 
DGLAP and MC computation schemes. For charged pions and 
protons+antiprotons we used experimental data from Refs.~\cite{expQCD}.
Below we shall present results of our SUSY MC simulations in terms of 
FFs for all pions $D_{\pi}$, all nucleons $D_N$ and all hadrons $D_h$.
We introduce the ratios $\eps_N(x)$ and $\eps_{\pi}(x)$ as:
$D_N(x) = \eps_N(x) D_h(x)$ and $ D_{\pi}(x) = \eps_{\pi}(x) D_h(x)$.
\unitlength1.0cm
\begin{figure}
\epsfig{file=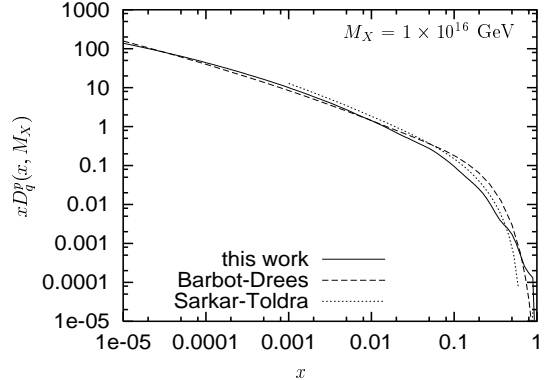,,height=5cm,width=7cm}
\caption{\label{fig2}
Comparison of nucleon spectra from present work computed with 
DGLAP equation (solid line), from Barbot \& Drees \cite{previous} 
(dashed line) and Sarkar \& Toldr\`a \cite{previous} (dotted line).
All three calculations are performed with quark as initial parton
for $M_X=1\times 10^{16}$ GeV.}
\end{figure}
The spectra of pions and nucleons at large $M_X$ have approximately the
same shape as the hadron spectra, and one can use in this case
$\eps_{\pi}=0.73\pm 0.03$ and $\eps_{N}=0.12\pm 0.02$ \cite{ABK},   
taking into account the errors in the experimental 
data \cite{expQCD}. We can calculate now the spectra of photons and neutrinos 
produced by the decays of pions neglecting the small contribution 
($0.15 \pm 0.04$) of $K$, $D$, $\Lambda$ and other particles. Including 
these particles affects stronger neutrinos than nucleons and photons, which 
are the main topic of this Section.

The normalized photon and neutrino spectrum from the decay of one $X$ 
particle at rest can be computed using the pion and nucleon FF following
the recipe given in \cite{ABK}.
We shall compare our photon spectra with those calculated by the DGLAP
method in \cite{previous}. The photon spectrum is
most interesting to compare, because it is straightforwardly
related to the hadron spectrum which is the basic physical quantity.
Moreover, the photon spectrum is the dominant component of radiation
produced by superheavy particles.

To be precise, we compare the FF $D_q^{\gamma,N,\nu}(x,M_X)$ at 
$M_X=1\times 10^{16}$ GeV. Figure \ref{fig2}, that is referred to nucleons, 
demonstrates good agreement between our spectrum and those from 
\cite{previous} at $x \leq 0.3$. 
As it was mentioned above, the disagreement at large $x$ is not surprising. 
Apart from $D_q^h(x,M_Z)$ taken directly from 
the experiments, both calculations use the much more uncertain
$D_g^h(x,Q^2)$. In our case, $D_g^h(x,Q^2)$ is taken from our MC
simulation~\cite{ABK}, in the case of Barbot \& Drees in \cite{previous} from
the fit performed in Ref.~\cite{Kniehl}. In both cases, rather large
uncertainties exist at large $x$ \cite{Kniehl}. 
The agreement between the three curves as presented in Figure \ref{fig2} 
is good. The same conclusion holds for the comparison of photons and neutrino 
spectra \cite{ABK}.
\unitlength1.0cm
\begin{figure}
\epsfig{file=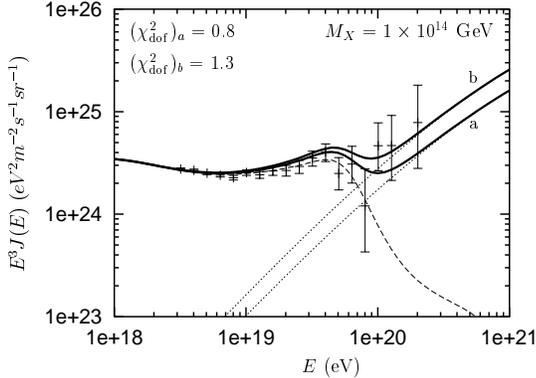,,height=5cm,width=7cm}
\caption{\label{fig3}
Comparison of SHDM prediction with the AGASA data. The calculated
spectrum of SHDM photons is shown by dotted curves for two different
normalizations. The dashed curve gives the spectrum of extragalactic
protons from uniformly distributed astrophysical sources. The sum of
these two spectra is shown by the thick curves. The $\chi^2$ values  
are given of the comparison of these 
curves with experimental data for $E\geq 4\times 10^{19}$~eV.}
\end{figure}
\section{UHECR from SuperHeavy particles and Topological Defects}

As follows from the previous section, the accuracy of spectrum 
calculations has reached such a level that one can consider the spectral
shape as a signature of the model. The predicted spectrum is
approximately $\propto dE/E^{1.9}$ in the region of $x$ at interest.
Another interesting feature of these new calculations is a decrease
of the ratio of photons to nucleons, $\gamma/N$, in the generation
spectrum. At $x\sim 1\times 10^{-3}$ this ratio is characterized by a value 
of 2 -- 3 only \cite{ABK}. The decrease of the $\gamma/N$ ratio is caused by 
a decrease of the number of pions in the new calculations. This result has 
an important impact for SHDM and topological defect models because the 
fraction of nucleons in the primary radiation increases. However, in both 
models photons dominate (i.e. their fraction becomes $\gsim 50\%$) at 
$E\gsim (7-8)\times 10^{19}$~eV. 
\unitlength1.0cm
\begin{figure}
\epsfig{file=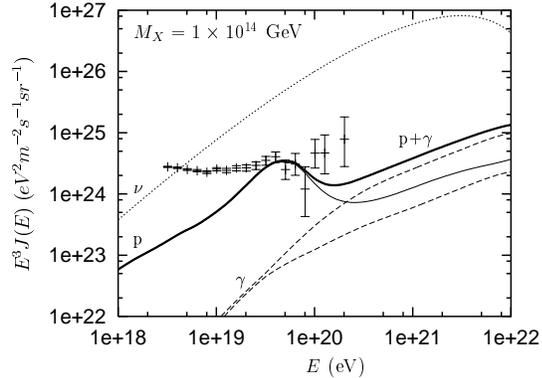,,height=5cm,width=7cm}
\caption{\label{fig4}
Diffuse spectra from necklaces. The upper curve shows neutrino flux,%
the middle - proton flux, and two lower curves - photon fluxes for
two cases of absorption. The thick continuous curve gives the sum
of the proton  and higher photon flux.}
\end{figure}
In this Section we shall consider: UHECR from superheavy dark matter
(SHDM) \cite{BKV97} and topological defects (TD) \cite{HiSch}.
Production of SHDM particles naturally occurs in the time-varying 
gravitational field of the expanding universe at the post-inflationary 
stage. The relic density of these particles is mainly determined
(at fixed reheating temperature and inflaton mass) by their
mass $M_X$. The range of practical interest is $(3 - 10)\times
10^{13}$ GeV, at larger masses the SHDM is a subdominant component of
the DM. SHDM is accumulated in the Galactic halo with the overdensity 
$\delta= \frac{\bar{\rho}_X^{\rm halo}}{\rho_X^{\rm extr}}=
\frac{\bar{\rho}_{\rm DM}^{\rm halo}}{\Omega_{\rm CDM}\rho_{\rm cr}}$,
where $ \bar{\rho}_{\rm DM}^{\rm halo}\approx 0.3$ GeV/cm$^3$, 
$\rho_{\rm cr}=1.88\times 10^{-29}h^2$ g/cm$^3$ and 
$\Omega_{\rm CDM}h^2=0.135$ \cite{WMAP}. With these numbers, 
$\delta \approx 2.1\times 10^5$. Because of this large overdensity,
UHECRs from SHDM have no GZK cutoff.

Clumpiness of SHDM in the halo can provide the observed small-angle
clustering.
The ratio $r_X=\Omega_X (t_0/\tau_X)$ of relic
abundance $\Omega_X$ and lifetime $\tau_X$ of the $X$ particle is
fixed by the observed UHECR flux as $r_X\sim 10^{-11}$. The numerical
value of $r_X$ is theoretically calculable as
soon as a specific particle physics and cosmological model is fixed. 
In the most interesting case of gravitational production of $X$
particles, their present abundance is determined by their
mass $M_X$ and the reheating temperature $T_R$.
Choosing a specific particle physics model one can fix also the life-time 
of the $X$ particle. There exist many models in which SH particles can
be quasi-stable with lifetime $\tau_X \gg 10^{10}$~yr.
The measurement of the UHECR flux, and thereby of $r_X$, selects from
the three-dimensional parameter space $(M_X, T_R, \tau_X)$ a
two-dimensional subspace compatible with the SHDM hypothesis.

In Figure \ref{fig3} we have performed a fit to the AGASA data using the 
photon flux from the SHDM model and the proton flux from uniformly distributed 
astrophysical sources. For the latter we have used the non-evolutionary model.
The photon flux is normalized to provide the best fit 
to the AGASA data at $E\geq 4\times 10^{19}$~eV. The fits are shown in 
Figure \ref{fig3} with $\chi^2$/d.o.f. indicated there. 
{\em One can see  from the fits in Figure \ref{fig3}, that the SHDM
model with the new spectra can explain only the excess of AGASA events 
at $E \gsim 1\times 10^{20}$~eV: depending on the SHDM spectrum
normalization and the details of the calculations for the extragalactic
protons, the flux from SHDM decays becomes dominant only  
above $(6-8)\times 10^{19}$~eV.} 

{\em Topological Defects\/} (for review see \cite{Xpart} and reference therein)
can naturally produce UHE particles. 
The following TD have been discussed as potential sources of UHE
particles: superconducting strings, ordinary strings,
monopolonium (bound monopole-antimonopole pair), monopolonia
(monopole-antimonopole pairs connected by a string), networks of 
monopoles connected by strings, vortons and necklaces (see Ref.~\cite{Xpart}
for a review and references).
Monopolonia and vortons are clustering in the Galactic halo and their 
observational signatures for UHECR are identical to SHDM. However the friction 
of monopolonia in cosmic plasma results in monopolonium lifetime much 
shorter than the age of the universe. 
Of all other TD which are not clustering in the Galactic halo, the most 
favorable for UHECR are {\em necklaces}. Their main phenomenological 
advantage is a small separation which ensures the arrival of highest
energy particles to our Galaxy. We shall calculate here the flux of UHECR 
from necklaces.

Necklaces are hybrid TD produced in the symmetry breaking pattern
$G \rightarrow H \times U(1) \rightarrow H \times Z_2$. At the first
symmetry breaking monopoles are produced, at the second one each
(anti-) monopole get attached to two strings. This system resembles
ordinary cosmic strings with monopoles playing the role of
beads. Necklaces exist as the long strings and loops.
The symmetry breaking scales of the two phase transitions, $\eta_m$
and $\eta_s$, are the main parameters of the necklaces. They determine
the monopole mass, $m \sim 4\pi \eta_m/e$, and the mass of the string per
unit length $\mu \sim 2\pi \eta_s^2$. The evolution of necklaces is
governed by the ratio $r\sim m/\mu d$, where $d$ is the average
separation of a monopole and antimonopole along the string. As it is
argued in Ref.~\cite{neckl}, necklaces evolve towards configuration with
$r\gg 1$. Monopoles and antimonopoles trapped in the necklaces inevitably 
annihilate in the end, producing heavy Higgs and gauge bosons ($X$ particles) 
and then hadrons. The rate of $X$ particles production in the universe can be
estimated as \cite{neckl} $\dot{n}_X \sim \frac{r^2 \mu}{t^3 M_X}$,
where $t$ is the cosmological time.

The photons and electrons from pion decays initiate e-m cascades and
the cascade energy density can be calculated as
$\omega_{\rm cas}=\frac{1}{2}f_{\pi}r^2\mu \int_0^{t_0} \frac{dt}{t^3}
\frac{1}{(1+z)^4}=\frac{3}{4}f_{\pi}r^2\frac{\mu}{t_0^2}$,
where $z$ is the redshift and $f_{\pi}\sim 1$ is the fraction of the 
total energy release transferred to the cascade. 
The parameters of the necklace model for UHECR are restricted by the 
EGRET observations~\cite{EGRET} of the diffuse gamma-ray flux.  
This flux is produced by UHE electrons and photons from
necklaces due to e-m cascades initiated in collisions with CMB photons. In the
range of the EGRET observations, $10^2 - 10^5$~MeV, the predicted spectrum
is $\propto E^{-\alpha}$ with $\alpha=2$ \cite{cascade}. The EGRET
observations determined the spectral index as $\alpha=2.10\pm 0.03$
and the energy density of radiation as $\omega_{\rm obs} \approx
4\times 10^{-6}$~eV/cm$^3$. The cascade limit consists in the 
bound $\omega_{\rm cas}\leq \omega_{\rm obs}$.

According to the recent calculations, the
Galactic contribution of gamma rays to the EGRET observations is
larger than estimated earlier, and the extragalactic gamma-ray
spectrum is not described by a power-law with $\alpha=2.1$. In this
case, the limit on the cascade radiation with $\alpha=2$ 
is more restrictive and is given by
$\omega_{\rm cas} \leq 2\times 10^{-6} {\rm eV/cm}^3$;
we shall use this limit in further estimates.
Using $\omega_{\rm cas}$ with $f_{\pi}= 1$ and $t_0=13.7$~Gyr \cite{WMAP}
we obtain from the limit on the cascade radiation 
$r^2\mu \leq 8.9\times 10^{27}$~GeV$^2$.

The important and unique feature of this TD is the small separation $D$
between necklaces. It is given by $D \sim r^{-1/2}t_0$ \cite{neckl}.
Since $r^2\mu$ is limited by e-m cascade radiation we
can obtain a lower limit on the separation between necklaces as
$D \sim \left (\frac{3f_{\pi}\mu}{4t_0^2\omega_{\rm cas}} \right )^{1/4}t_0
> 10(\mu/10^6~{\rm GeV}^2)^{1/4}~{\rm kpc}$,
this small distance is a unique property of necklaces allowing the
unabsorbed arrival of particles with the highest energies.
The fluxes of UHECR from necklaces are shown in Figure \ref{fig4}.
We used in the calculations $r^2\mu = 4.7\times 10^{27}$~GeV$^2$ which 
corresponds to $\omega_{\rm cas}= 1.1\times 10^{-6}$~ eV/cm$^3$, i.e. 
twice less than allowed by the bound on $\omega_{\rm cas}$. The mass of the $X$
particles produced by monopole-antimonopole annihilations is taken as  
$M_X= 1\times 10^{14}$~GeV. 
>From  Figure \ref{fig4} one can see that the necklace model for UHECR can 
explain only the highest energy part of the spectrum, with the AGASA excess 
somewhat above the prediction. This is the direct consequence of the new
spectrum of particles in $X$ decays obtained in this work. Thus UHE 
particles from necklaces can serve only as an additional component in
the observed UHECR flux. 

\section{Conclusions}

In this paper we have compared the MC and DGLAP methods for the
calculation of hadron spectra produced by the decay (or annihilation)
of superheavy $X$ particles with masses up to $M_{\rm GUT} \sim
1\times 10^{16}$~GeV. We found an excellent agreement of these two methods.
The calculations have been performed both for ordinary QCD and SUSY QCD. 
The inclusion of SUSY partons in the development of the cascade 
results only in small corrections, and it justifies our computation scheme 
with a single mass scale $M_{\rm SUSY}$ \cite{ABK}.

In comparison to the DGLAP method, the MC simulation has the advantage of
including coherent branching. It allows reliable calculations 
at very small $x$. The Gaussian peak, the signature of the QCD spectrum,
cannot be obtained using the DGLAP equations. We have calculated the 
all-hadron spectra, as well as spectra of charged pions and nucleons, using 
the DGLAP equations. 

Our nucleon spectrum agrees well with that of Refs.~\cite{previous}.
We compared also our spectrum of photons with the calculations 
of Ref. \cite{previous}. The comparison of the photon spectra is
interesting, because of physical reasons (photons can be observable
particles), and because the photon spectra are connected directly with the
hadron spectra. The spectra are in good agreement. 
We conclude that all calculations are in a good agreement especially
in the most interesting low $x$ regime and the predicted shape of the 
generation spectrum ($\propto dE/E^{1.9}$) can be considered as a signature 
of models with decaying (annihilating) superheavy particles.

The predicted spectrum of SHDM model cannot fit the observed UHECR
spectrum at $1\times 10^{18}~{\rm eV}\leq E \leq (6-8)\times 10^{19}$~eV.
Only events at $E\gsim (6-8)\times 10^{19}$~eV, and most notably the AGASA 
excess at these energies, can be explained in this model. The robust 
prediction of this model is photon dominance. In present calculations this 
excess diminishes to $\gamma/N \simeq 2 - 3$ \cite{ABK}. 
According to the recent calculations the muon content
of photon induced EAS at $E>1\times 10^{20}$ eV is high, but lower by
a factor 5 -- 10 than in hadronic showers. The muon content of EAS at
$E>1\times 10^{20}$~eV has been recently measured in
AGASA~\cite{expCR}. The measured value
is the muon density at the distance 1000~m from the shower core,
$\rho_{\mu}(1000)$. From 11 events at $E>1\times 10^{20}$~eV 
the muon density was measured in 6. In two of them with
energies about $1\times 10^{20}$~eV, $\rho_{\mu}$ is almost twice
higher than predicted for gamma-induced EAS. Taking into account the
contribution of extragalactic protons at this energy, the ratio $\gamma/p$ 
predicted by the SHDM model is 1.2 -- 1.4. It is lower than  
the upper limit $\gamma/p \leq 2$ obtained by AGASA 
at $E=3\times 10^{19}$~eV on the basis of a much larger
statistics. The muon content of the remaining 4 EAS  
marginally agrees with that predicted for gamma-induced showers. The 
contribution of extragalactic protons for these events is negligible,
and the fraction of protons in the total flux can be estimated as 
$0.25 \leq p/{\rm tot}\leq 0.33$. This fraction gives a considerable
contribution to the probability of observing 4 showers with slightly
increased muon content. Not excluding the SHDM model, the AGASA events
give no evidence in favor of it.

The simultaneous observation of UHECR events in fluorescent light and 
with water Cherenkov detectors has a great potential to distinguish 
between photon and proton induced EAS. An anisotropy towards the
direction of the Galactic Center is another signature of the SHDM model. 
Both kinds of informations from Auger~\cite{expCR} will be
crucial for the SHDM model and other top-down scenarios.   
Topological defect models are another case when short-lived superheavy
particle decays can produce UHECR. In Figure \ref{fig4} the spectra from
necklaces are presented. One can see that at $E \gsim 1\times 10^{20}$~eV
photons dominate, and the discussion in the previous paragraph applies
here too. In contrast to previous calculations,
the agreement with observations is worse: necklaces can explain only the 
highest energy part of the spectrum in Figure \ref{fig4}, with the AGASA 
excess somewhat above the prediction. In the other energy ranges, 
UHE particles from necklaces can provide only a subdominant
component. Other TDs suffer even more problems.

\end{document}